\documentclass[twocolumn,english,pra,showpacs,superscriptaddress,longbibliography]{revtex4-1}
\usepackage{amsfonts}
\usepackage{amsmath}
\usepackage{amssymb}
\usepackage[colorlinks, citecolor=blue,linkcolor=red]{hyperref}
\usepackage{color}
\usepackage{float}
\usepackage{hyperref}
\usepackage{textcomp}
\usepackage[rightcaption]{sidecap}
\usepackage{subfigure}
\usepackage[rightcaption]{sidecap}
\usepackage{graphicx}
\begin{document}
\title{Spin-Orbit Coupling Induced Back-action Cooling in Cavity-Optomechanics with a Bose-Einstein Condensate}
\author{Kashif Ammar Yasir}
\email{kayasir@iphy.ac.cn}\affiliation{Beijing National Laboratory for Condensed Matter Physics, Institute of Physics, Chinese Academy of Sciences, Beijing 100190, China.}
\affiliation{School of Physical Sciences, University of Chinese Academy of Sciences, Beijing 100190, China.}
\author{Lin Zhuang}
\email{stszhl@mail.sysu.edu.cn}\affiliation{School of Physics, Sun Yat-Sen University, Guangzhou 510275, P. R. China.}
%\email{kashif\_ammar@yahoo.com}
\author{Wu-Ming Liu}
%\email{wliu@iphy.ac.cn}
\email{wliu@iphy.ac.cn}\affiliation{Beijing National Laboratory for Condensed Matter Physics, Institute of Physics, Chinese Academy of Sciences, Beijing 100190, China.}
\affiliation{School of Physical Sciences, University of Chinese Academy of Sciences, Beijing 100190, China.}
%%%%%%%%%%%%%%%%%%%%%%%%%%%%%%%%%%%%%%%%%%%%%%%%%%%%%%%%%%%%%%%%%%%%%%%%%%%%%%%%%%%%%%%%%%%%%%%%%%%%%%%%%%%%%%%%%%%%%%%%%%%

\begin{abstract}
We report a spin-orbit coupling induced back-action cooling in an optomechanical system, composed of a spin-orbit coupled Bose-Einstein condensate 
trapped in an optical cavity with one movable end mirror, by suppressing heating effects of quantum noises. 
The collective density excitations of the spin-orbit coupling mediated hyperfine states -- serving as atomic oscillators equally coupled to the cavity 
field -- trigger strongly driven atomic back-action. We find that the back-action not only revamps low-temperature dynamics of its own but also provides an opportunity to cool the mechanical mirror to its quantum mechanical 
ground state. Further, we demonstrate that the strength of spin-orbit coupling also superintends dynamic structure factor and squeezes nonlinear 
quantum noises, like thermo-mechanical and photon shot noise, which enhances optomechanical features of 
hybrid cavity beyond the previous investigations.
Our findings are testable in a realistic setup and enhance the functionality of cavity-optomechanics with spin-orbit coupled hyperfine 
states in the field of quantum optics and 
quantum computation.
\end{abstract}
%\date{\today}
\maketitle

\section{INTRODUCTION}
Cavity-optomechanics provides splendid foundations in utilizing mechanical effects of light to couple optical degree of freedom 
with mechanical degree of freedom \cite{Kippenberg08,Kipp09,Meystre13}. 
A pivotal paradigm was to cool vibrational 
modes of mechanical degree of freedom to its quantum mechanical ground state which has been attempted to achieve via 
laser radiations, electronic feedback and dynamical back-action \cite{CornnellNat2010,teufel2011,chan2011,Steele2015,Arcizet2006,Gigan2006,Kleckner2006,Kippenberg2006,Teufel2006}. 
The dynamical back-action is the cavity delay induced by the interactions of intra-cavity radiation pressure and the Brownian 
motion of the mirror which leads to cool mirror depending upon detuning \cite{wilson2015,Wilson2007,Wollman2015,nigues2015}. 
The demonstration of cavity-optomechanics with other physical objects, like cold atomic gases \cite{Murch2008} and Bose-Einstein condensate (BEC) \cite{Esslinger,kashif1,kashif2}, 
opens up various new aspects to further cool vibrational modes through atomic back-action \cite{peter2,Pater06,Sonam2013,liu2013,Elste2009}. 
However, thermo-mechanical heating, due to the quantum noises \cite{zhang20141,zhang20142,weber2016,amir,Clerk}, is a major obstacle in achieving an oscillator with long phonon lifetime 
in quantum ground state which we intend to solve by the inclusion of spin-orbit (SO)-coupled BEC. 

The SO-coupling, a stunning phenomenon describing interaction between spin of quantum particle and its momentum, has made remarkable 
progress in the last few years \cite{Galitski2013,Liu2009,Lin2009} and is essential to analyze spin-Hall effect \cite{Kato2004,Konig2007}, topological insulators \cite{Bernevig2006,Hsieh2008,Li2013,Sun2012} 
and spintronic devices \cite{Koralek2009}. 
The demonstration of SO-coupling in one-dimensional optical lattices \cite{Hamner2015,Lin2011,Hu2012,Jiang2014,Zhou2015,Cai2012,liu1,xiang2014,Goldman2014,Dalibard2011,wu2015,wu2011} and 
optical cavities \cite{Deng2014,Padhi2014,Dong2014} enables us to utilize this phenomenon in optomechanical environment. 
The SO-coupling induces significant variations in cavity radiation pressure by separating atomic mode in hyperfine spin-states 
which gives rise to self-confinement via dynamical back-action \cite{Kippenberg08,pu2015,dennett2016,Wilson2008}.
Further, dynamic structure factor, 
a phenomenon describing density fluctuations and mean energy of excited quasi-particles, is crucially important in quantum many-body systems \cite{Punk2014} 
and provides explanation of quasi-particle evolution under noise effects \cite{Miyake2011,Sachdev2000,Landig2015}.
\begin{figure}[bp]
\includegraphics[width=9cm]{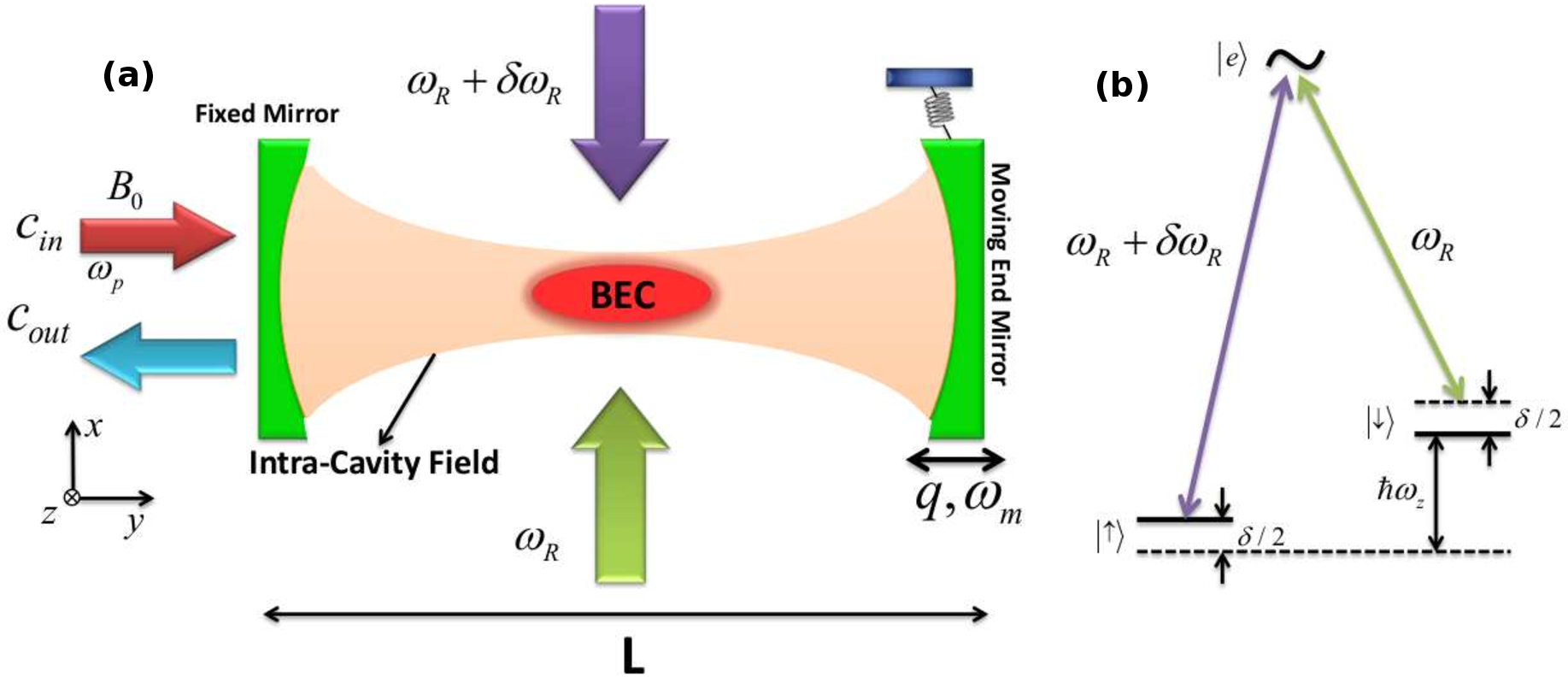}
\caption{(Color online) (a) Schematic diagram of spin-orbit (SO)-coupled $^{87}Rb$ Bose-Einstein condensate (BEC) trapped inside 
a high-\textit{Q} Fabry-P\'{e}rot cavity with one moving end mirror, where $\hat{y}$-axis is the cavity axis and $\hat{x}$-axis is the direction of incident Raman beams. (b) Energy level configuration of SO-coupled BEC.}
\label{fig11}
\end{figure}

In this paper, we report the SO-coupling induced back-action cooling mechanism in a hybrid optomechanical cavity with SO-coupled BEC and 
one moving end mirror in the presence of quantum noises. 
We show that the SO-coupling induced features modify intra-cavity atomic back-action which not only leads to maneuver low-temperature 
dynamics of atomic mode but also helps in ground state cooling of vibrational modes of the cavity mirror.
Further, the coupling of mechanical mirror with cavity modifies the eigenenergy spectrum of hyperfine states via transformation of phonons to 
atomic degree of freedom and provides a way to tune quantum phase transitions in BEC.
Furthermore, we compute dynamic structure factor by manipulating two frequency auto-correlation of photons leaking-out from cavity and observe the influence of 
SO-coupling on dynamic structure factor.

\section{CAVITY-OPTOMECHANICS WITH SO-COUPLED BEC}
The system consists of a high-\textit{Q} Fabry-P\'{e}rot cavity, with one fixed and one movable mirror, containing SO-coupled BEC illuminated along $\hat{y}$-axis and coherently driven by 
single-mode optical field with frequency $\omega_{p}=\omega_R+\delta\omega_R=8.8\times2\pi$ MHz, see Fig.\ref{fig11}(a).  
To produce SO-coupling, we chose two internal atomic pseudo-spin-states 
in $N=1.8\times10^5$ $^{87}Rb$ bosonic 
particles having $F=1$ electronic ground manifold of $5S_{1/2}$ electronic levels labeled as 
$|\uparrow\rangle =| F=1, m_F=0\rangle$ (pseudo-spin-up) and $|\downarrow\rangle=| F=1, m_F=-1\rangle$ (pseudo-spin-down), 
as shown in Fig.\ref{fig11}(b). The magnetic $10 G$ bias field $B_0$ is 
applied along cavity axis ($\hat{y}$-axis) to induce Zeeman shift $\hbar\omega_z$, where $\omega_z\approx4.8\times2\pi$ kHz. 
Two counter-propagating Raman lasers along $\hat{x}$-axis, with wavelength $\lambda=804.1 nm$ 
and detuning $\delta=1.6E_R$, interact with atomic spin-states in opposite direction. The frequencies of these Raman beams are 
$\omega_R$ and $\omega_R+\delta\omega_R$, respectively, with constant frequency difference $\delta\omega_R=\omega_z+\delta/\hbar\backsimeq4.8\times2\pi$ MHz. 
$\pmb{k_L}=\hbar k_y=\sqrt{2}\pi\hbar/\lambda$ and $E_R=(\hbar k_y)^2/2m_a=20\times2\pi$ kHz represent unit-less momentum and energy, respectively. 
The mechanical mirror is coupled 
to the cavity mode, oscillating 
with frequency $\omega_{c}=4\times2\pi MHz$ and detuning $\Delta_c=\omega_p-\omega_c=\delta\omega_R$, 
via radiation pressure force \cite{Esslinger,Kippenberg08}. 

The system Hamiltonian consists of three parts, $\hat{H}=\hat{H}_{a}+\hat{H}_{m}+\hat{H}_{f}$. 
In strong detuning regime and under rotating-wave approximation, the many-body Hamiltonian for atomic mode ($H_{a}$) 
is given as \cite{Lin2011,Deng2014,Padhi2014},
\begin{eqnarray}\label{Ha1}
\hat{H}_a &=& \int dr\pmb{\hat{\psi}^{\dag}}(\pmb{r})\bigg(H_{0}+V_{LAT}\bigg) \pmb{\hat{\psi}}(\pmb{r})\nonumber\\
&+&\frac{1}{2}\int d\pmb{r}\sum_{\sigma,\acute{\sigma}} U_{\sigma,\acute{\sigma}}\hat{\psi}^{\dag}_\sigma(r)\hat{\psi}^{\dag}_{\acute{\sigma}}(r)\hat{\psi}_{\acute{\sigma}}(r)\hat{\psi}_\sigma(\pmb{r}),
\end{eqnarray}
where $m_a$ is the mass of an atom, $\pmb{\hat{\psi}}=[\hat{\psi}_\uparrow,\hat{\psi}_\downarrow]^{T}$ represents bosonic field operator for 
pseudo-spin-up and -down atomic states. $ H_{0}=\hbar^2\pmb{k}^2\sigma_{0}/2m_a+\tilde{\alpha}\pmb{k_x}\sigma_{y}+\frac{\delta}{2}\sigma_y+\frac{\Omega_z}{2}\sigma_z$ 
describes single-particle Hamiltonian containing SO-coupling terms \cite{Lin2009,Fetter2014}, where 
$\tilde{\alpha}=E_R/\pmb{k_L}$ is the strength of SO-coupling. $\delta=-g\mu_B B_z$ 
and $\Omega_z=-g\mu_B B_y$ are related to the Zeeman field effects along $\hat{z}$ and $\hat{y}$ axis, respectively. 
$\sigma_{x,y,z}$ represents $2\times2$ Pauli matrices under pseudo-spin rotation and $\sigma_0$ is a unit matrix \cite{Lin2011,Fetter2014}. 
$V_{LAT}=\hbar \hat{c}^{\dag}\hat{c}U_0[cos^2(kx)+cos^2(ky)]$ is the intra-cavity 
optical lattice under assumption $k_x=k_y=k$ and both atomic states are equally coupled to the cavity because of having same motional quantum state \cite{Padhi2014,ritsch}. $\hbar \hat{c}^{\dag}\hat{c}U_0$ 
is optical potential depth with atom-photon coupling $U_0=g_0^2/\Delta_a$, where $g_{0}$ is the vacuum Rabi frequency and $\Delta_{a}$ 
is far-off detuning between field frequency and atomic transition frequency $\omega_{0}$. Here, $\hat{c} (\hat{c}^\dag)$ are annihilation (creation) operators for cavity mode. Finally, last term explains 
many-body intra-species and inter-species interactions for atomic spin-states, where $\sigma,\acute{\sigma}\in\{\uparrow,\downarrow\}$. 
$U_{\sigma,\acute{\sigma}}=4\pi a_{\sigma,\acute{\sigma}}^2\hbar^2/m_a$ accounts for strength of atom-atom interactions, where $a_{\sigma,\acute{\sigma}}$ is the s-wave 
scattering length.

The Hamiltonian for moving end mirror is $\hat{H}_{m}=\hbar\omega_{m}\hat{b}^{\dag}\hat{b}-
i\hbar\frac{ g_m}{\sqrt{2}}\hat{c}^{\dag}\hat{c}(\hat{b}^{\dag}+\hat{b})$, where first term describes the motion of mechanical mirror with frequency $\omega_m$ 
and $\hat{b} (\hat{b}^\dag)$ are annihilation (creation) operators for mechanical mirror with commutation relation 
$[\hat{b}^\dag,\hat{b}]=1$. Second term accommodates mechanical mirror coupling with cavity mode with coupling strength 
$g_m=\sqrt{2}(\omega_{c}/L)x_{0}$, where  
$x_{0}=\sqrt{\hbar/2m\omega_{m}}$ is zero point motion of mechanical mirror having mass $m$. $\hat{H}_{c}=\hbar\triangle_{c}\hat{c}^{\dag}\hat{c}-i\hbar\eta(\hat{c}
-\hat{c}^{\dag})$, where first term is the strength of cavity mode and second part is associated with 
its coupling with external pump field with strength $\vert\eta\vert=\sqrt{P\kappa/\hbar\omega_{p}}$, where $P$ is the input field power.

We substitute plane-wave ansatz 
$\pmb{\hat{\psi}}(r)=e^{ikr}\pmb{\hat{\varphi}}$, where $\pmb{\hat{\varphi}}=[\hat{\varphi}_\uparrow,\hat{\varphi}_\downarrow]^{T}$, 
in atomic mode Hamiltonian $H_a$ by considering homogeneous atomic modes distribution with normalization condition 
$|\hat{\varphi}_\uparrow|^2+|\hat{\varphi}_\downarrow|^2=N$. We assume that the strengths of intra-species interactions 
of both spin-states are equal with each other and are defined as, 
$U_{\uparrow,\uparrow}=U_{\downarrow,\downarrow}=U$. Similarly, inter-species interactions can be modeled as, 
$U_{\uparrow,\downarrow}=U_{\downarrow,\uparrow}=\varepsilon U$, where parameter $\varepsilon$ depends upon the incident laser 
configuration \cite{Lin2009}. Under these considerations, we solve equation $H_a$ and compute quantum Langevin equations for the system by using standard quantum-noise operators to include quantum noises and dissipations associated with the system \cite{Dalibard2011,Pater06}. 
The quantum Langevin equation helps us in developing coupled and time dependent set of equations, containing 
noise operators for optical, mechanical and atomic degrees of freedom,
\begin{center}
	\begin{eqnarray}\label{2}
	\frac{d\hat{c}}{dt}&=&\dot{\hat c}=(i\tilde{\Delta}+i\frac{g_m}{\sqrt{2}}(\hat{b}+\hat{b}^\dag)
	-ig_{a} \pmb{\hat{\varphi}^\dag\hat{\varphi}}-\kappa)\hat{c}+\eta\nonumber\\
	&&+\sqrt{2\kappa} a_{in},\\
	\frac{d\hat{b}}{dt}&=&\dot{\hat b}=-\omega_{m}\hat{b}-\frac{g_m}{\sqrt{2}}\hat{c}^{\dag}\hat{c}
	-\gamma_{m}\hat{b}+\sqrt{\gamma_{m}}f_m, \\
	\frac{d\pmb{\hat{\varphi}}}{dt} &=&\dot{\pmb{\hat{\varphi}}} =(\frac{\hbar\pmb{k}^2\sigma_{0}}{2m}+\tilde{\alpha}\pmb{k_x}\sigma_{y}+
	\frac{\delta}{2}\sigma_y+\frac{\Omega_z}{2}\sigma_z-\gamma_a+g_a\hat{c}^{\dag}\hat{c})\pmb{\hat{\varphi}}\nonumber\\
	&&+\frac{1}{2}U\pmb{\hat{\varphi}}^\dag\pmb{\hat{\varphi}}\pmb{\hat{\varphi}}
	+\frac{1}{2}\varepsilon U \hat{\varphi}_\sigma^\dag\hat{\varphi}_{\acute{\sigma}}\hat{\varphi}_\sigma++\sqrt{\gamma_{a}}f_a,
	\end{eqnarray}
\end{center}
where $\tilde{\Delta}=\Delta _{c}-NU_{0}/2$ is the modified detuning of the system and $\hat{c}_{\mathrm{in}}$ is Markovian 
input noise operator associated with intra-cavity field, having zero-average $\langle \hat{c}_{in}(t)\rangle=0$ 
and delta-correlation $\langle \hat{c}_{in}(t)\hat{c}_{in}^{\dagger}(\acute{t})\rangle=\delta(t-\acute{t})$ under the condition $\hbar\omega_c>>k_BT$. 
The term $\gamma _{m}$ describes mechanical energy decay rate of the moving end mirror and 
$\hat{f}_{m}$ is noise operator (or zero-mean Langevin-force operator) connected with the Brownian motion of 
mechanical mirror and can be defined by using non-Markovian correlation \cite{Dalibard2011,Pater06} 
$\langle \hat{f}_m(t)\hat{f}_m(\acute{t})\rangle=\frac{\gamma_m}{2\pi\omega_m}\int d\omega e^{-i\omega(t-\acute{t})}[1+Coth(\frac{\hbar\omega}{2k_BT})]$. 
The external harmonic trapping potential of the condensate, which
we have ignored so far because it appeared to be spin independent, cause the damping of the atomic motion.
The parameter $\gamma_{a}$ represents such damping of atomic dressed states motion while $\hat{f}_{a}$ 
is the associated noise operators assumed to be Markovian with delta-correlation 
$\langle \hat{f}_{a}(t)\hat{f}_{a}^{\dagger}(\acute{t})\rangle=\delta(t-\acute{t})$ under the condition $\hbar\Omega>>k_BT$. 
Further, $\sigma,\acute{\sigma}\in\{\uparrow,\downarrow\}$ and $g_{a}=\frac{\omega_{c}}{L}\sqrt{\hbar/m_{bec}4\omega_{r}}$ is 
the coupling of atomic mode with intra-cavity field, having effective BEC frequency $\Omega=4\omega_r$ and mass 
$m_{bec}=\hslash\omega_{c}^{2}/(L^{2}U^2_{0}\omega_{r})$, where $\omega_r=3.8\times2\pi$kHz is the recoil frequency of atoms and $L=1.25\times10^{-4}$m is the cavity length.

\begin{figure}[tp]
	\includegraphics[width=8.5cm]{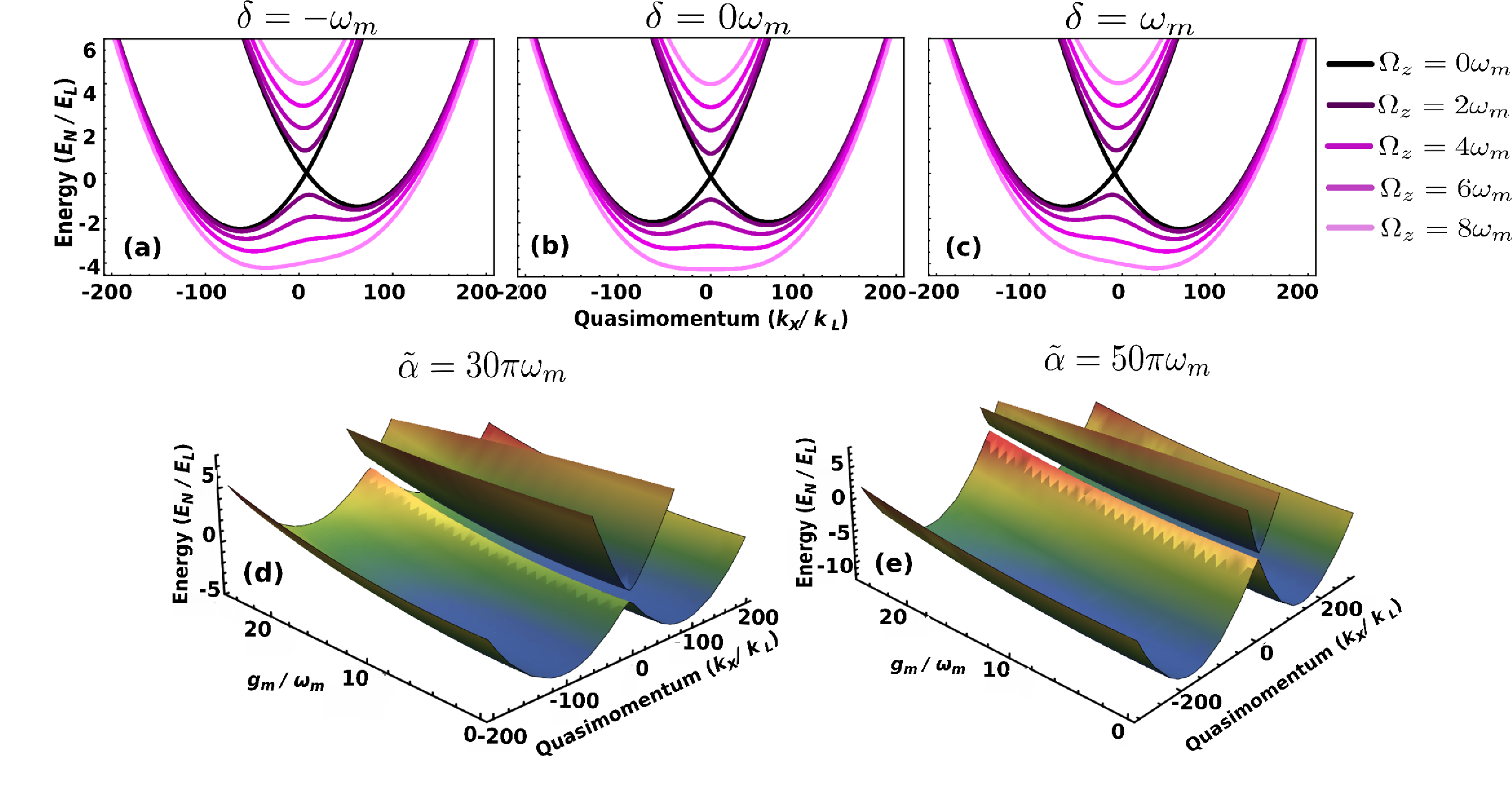}
	\caption{(Color online) (\textbf{a}-\textbf{c}) Eigenenergies spectrum $E_N$ of spin-orbit (SO)-coupled Bose-Einstein condensate (BEC) 
		as a function of quasi-momentum $\pmb{k_x/k_L}$ for different 
		$\Omega_z/\omega_m$ and $\delta/\omega_m$, when 
		$g_m/\omega_m=0.1$ and $\alpha/\omega_m=20\pi$. (It should be noted that we consider $\pmb{k_y}=\pmb{k_z}=0$ because SO-coupling is occurring only in the direction of 
		$\hat{x}$-axis.) The black curve represents dispersion at $\Omega_z/\omega_m=0$ while magenta shaded curves (from darkest to lightest) correspond to 
		$\Omega_z/\omega_m=2, 4, 6, 8$, respectively. 
		(\textbf{a}), (\textbf{b}) and (\textbf{c}) show the behavior of dispersion $E_N$ for 
		$\delta/\omega_m=-1, 0, 1$, respectively. (\textbf{d}) and (\textbf{e}) show dispersion 
		$E_N$ versus $\pmb{k_x/k_L}$ and $g_m/\omega_m$, with 
		$\alpha/\omega_m=30\pi$ and $\alpha/\omega_m=50\pi$, respectively, at $\Omega_z=\omega_m$ and $\delta/\omega_m=0$. 
		The other parameters used are $U/\omega_m=5.5$, 
		$\varepsilon/\omega_m=0.1$, $\kappa/\omega_m=0.1$, $\gamma_a/\omega_m=0.01$, $\gamma_m/\omega_m=0.05$ and mechanical mirror frequency $\omega_m\approx\omega_c-\omega_p$.}
	\label{fig1}
\end{figure}
\section{CONDENSATE DISPERSION SPECTRUM}
The eigenenergy spectrum is calculated from time-dependent quantum Langevin equations (for details see Appendix \ref{ap1}). The 
SO-coupling will create two minima corresponding to lowest energy-levels of atomic spin-states, as illustrated in Fig.\ref{fig1}. 
At $\Omega_z/\omega_m=0$, no band-gap appears between lower and upper dispersion states causing the 
phase mixing of atomic dressed states. However, in presence of Raman coupling, the band-gap 
between upper $E_N>0$ and lower $E_N<0$ dispersion states appears in the form of Dirac-cone which increases with the increase in Raman coupling. The higher values of $\Omega_z/\omega_m$ merge two minima corresponding to the dressed states into single minima causing quantum phase transitions 
from mixed phase to separate phase of atomic mode. 
It can also be seen that the non-zero Raman detuning $\delta/\omega_m\neq0$ leads to the 
symmetry-breaking of dispersion over quasi-momentum. For the small value of Raman coupling ($\Omega_z<4\omega_m$), the dispersion appears in the form of double-well potential in the quasi-momentum which leads to the zero group velocity of atoms \cite{HI2004}. The asymmetric behavior indicates 
rapid population transfer and enhancement in band-gap induced features of hyperfine states in cavity environment. 
Moreover, it is noted that because of cavity confinement, the atomic quasi-momentum $\pmb{k_x/k_L}$ interacts with the optical mode along cavity axises. Thus, SO-coupled BEC face an anisotropic potential which leads towards spatial spread of BEC energy spectrum 
along cavity axis, as can be seen in Fig.\ref{fig1}
 
The coupling between atomic states 
and intra-cavity potential is disturbed by the existence of mechanical mirror, and vice-versa, when 
the atomic modes become resonant with the optical sideband. At this point, atomic spin states will absorb some phonons
emitted by the mechanical mirror via cavity mode 
and will behave as a phononic-well. Therefore, the increase in mirror-field coupling gives rise to 
atomic-state energy levels, as shown in Fig.\ref{fig1}(d) and \ref{fig1}(e), which provides precise control over the dispersion relation of atomic energy spectrum and quantum 
phase transitions of BEC.

\section{ATOMIC DENSITY-NOISE SPECTRUM}
\begin{figure}[htp]
	\includegraphics[width=8.5cm]{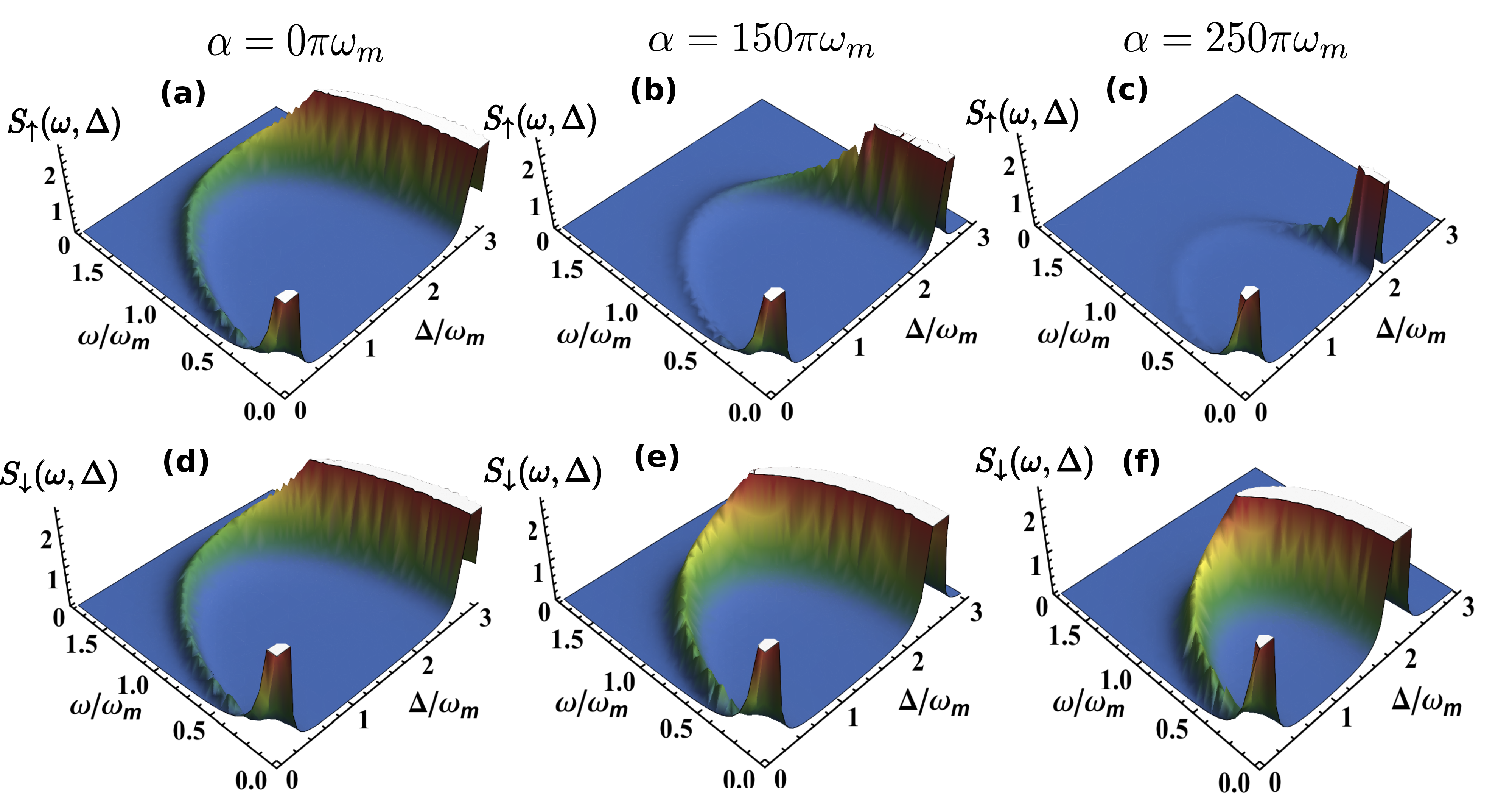}
	\caption{(Color online)(\textbf{a-c}) Illustrate DNS $S_\uparrow(\omega,\Delta)$ as a function of $\Delta/\omega_m$ and $\omega/\omega_m$ for 
		$\alpha/\omega_m=0\pi, 150\pi$ and $250\pi$, respectively, when $\Omega_z=\omega_m$, $\delta/\omega_m=1$ and $G_a/\omega_m=28.5$. 
		(Note: The color configuration corresponds to the strength 
		of DNS.) Similarly, (\textbf{d-f}) demonstrate DNS $S_\downarrow(\omega,\Delta)$ for $\alpha/\omega_m=0\pi, 150\pi$ and $250\pi$, respectively. 
		Here $G_m/\omega_m=1.5$, $\Omega/\omega_m=70.8$, $\omega_m=3.8\times2\pi$kHz and the thermal reservoir temperature is taken as $T=300$K.}\label{fig2}
\end{figure}
We calculate density-noise spectrum (DNS) of atomic spin states by taking two-frequency auto-correlation of the frequency domain solution of quantum Langevin 
equations, 
$S_{\uparrow,\downarrow}(\omega,\Delta)=\frac{1}{4\pi}\int e^{-i(\omega+\omega^{\prime})t}\langle \delta\hat{q}_{\uparrow,\downarrow}(\omega)\delta\hat{q}_{\uparrow,\downarrow}(\omega^{\prime})+\delta\hat{q}_{\uparrow,\downarrow}(\omega^{\prime})\delta\hat{q}_{\uparrow,\downarrow}(\omega)\rangle d\omega^{\prime}$, 
where $\delta\hat{q}_{\uparrow,\downarrow}$ are dimensionless position quadratures of spin states defined as, $\delta\hat{q}_{\uparrow,\downarrow}=\frac{1}{\sqrt{2}}(\hat{\varphi}_{\uparrow,\downarrow}+\hat{\varphi}_{\uparrow,\downarrow}^{\dag})$. Here the effective system detuning is $\Delta = \tilde{\Delta}-g_m q_s +g_{a}N$, where $q_s$ is steady-state position quadratures of mechanical mirror, 
while $G_{m}=\sqrt{2}g_m|c_{s}|$ and $G_{a}=\sqrt{2}g_{a}|c_{s}|$ are the effective couplings of intra-cavity field with mechanical mirror and atomic modes, respectively, 
tuned by the steady-state cavity mode amplitude $c_{s}=\frac{\eta}{\kappa +i\Delta}$ (for detailed calculations see Appendix \ref{ap2} and \ref{ap3}). By considering the correlation 
operators of Markovian and Brownian noises in frequency domain \cite{peter2,Pater06,f2005,Pirkkalainen2015}, we plot the DNS for pseudo spin-$\uparrow$ and 
spin-$\downarrow$ atomic states as shown in Fig.\ref{fig2}. 

The inclusion of SO-coupling in trapped-atoms modifies atomic back-action generated by the interaction of intra-cavity radiation pressure 
with BEC. These modifications enhance the cavity induced self-regulatory mechanism of atomic mode. 
In the absence of SO-coupling ($\alpha/\omega_m=0\pi$), both $S_\uparrow(\omega,\Delta)$ and $S_\downarrow(\omega,\Delta)$ behave in a similar way as shown 
in Fig.\ref{fig2}(a) and \ref{fig2}(d) \cite{peter2,Pater06}, where $\alpha=\tilde{\alpha}\pmb{k_x}$ is the effective strength of SO-coupling. 
Both the cooling as well as heating mechanisms are observable 
because area under $S_\uparrow(\omega,\Delta)$ describes the effective temperature of atomic mode, as shown in effective temperature calculation of mechanical mirror in next section. One can observe 
a semi-circular structure appearing with the increase in $\Delta/\omega_m$ caused by the red-shift 
in the peak frequency of $S_\uparrow(\omega,\Delta)$. 
Height of the structure initially decreases with increase in $\Delta/\omega_m$ towards $\omega/\omega_m$ but 
shortly again starts rising along the semi-circular structure. 
The optimal cooling is achieved at $\Delta=\omega_m/2$ with a considerable shrink in the area underneath atomic DNS. 

In the presence of $\alpha/\omega_m$, $S_\uparrow(\omega,\Delta)$ and $S_\downarrow(\omega,\Delta)$ start behaving in 
a different manner because of the emergence of Zeeman shift among the hyperfine atomic states with SO-coupling. For $S_\uparrow(\omega,\Delta)$, 
the height of semi-circular 
structure is suppressed due to the energy transformation via intra-cavity potential, as 
can be seen in Fig.\ref{fig2}(b) and \ref{fig2}(c), where $\alpha/\omega_m=150\pi$ and $250\pi$, respectively. 
The optimal cooling point is now shifted to $\Delta/\omega_m=1$. 
The existence of SO-coupling not only decreases the area underneath $S_\uparrow(\omega,\Delta)$ but also suppresses the 
radius of that semi-circular structure providing controlled cooling of atomic mode. 
\begin{figure}[tp]
\centering
\includegraphics[width=8.5cm]{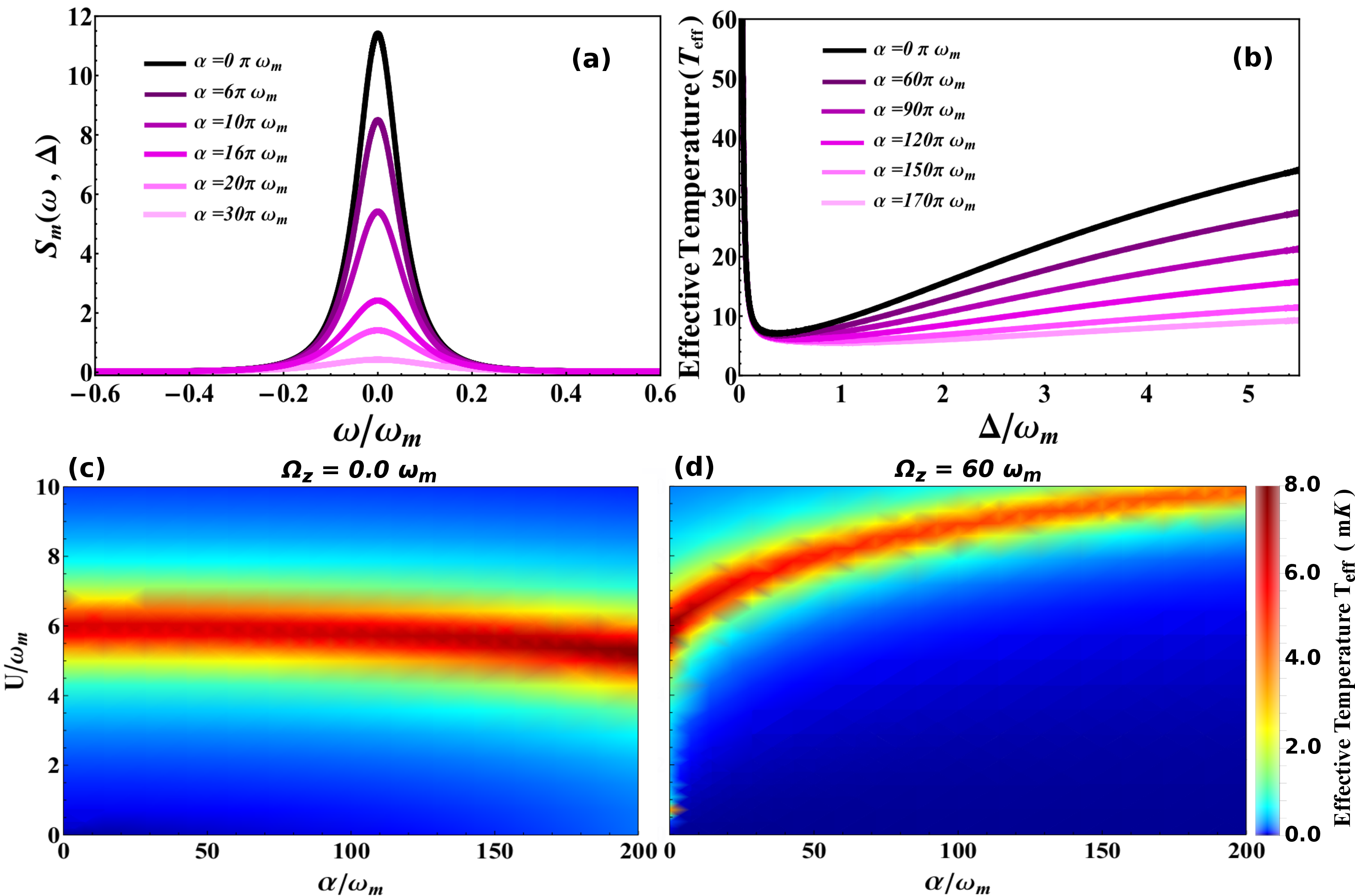}
\caption{(Color online) (\textbf{a}) Illustrates $S_m(\omega,\Delta)$ versus normalized frequency $\omega/\omega_m$ at 
$\Delta/\omega_m=1.8$, $G_m/\omega_m=10$ and $G_a/\omega_m=20$. The black curve is for $\alpha/\omega_m=0\pi$ and magenta curves from dark shade to light shade represent $\alpha/\omega_m=6\pi, 
10\pi, 16\pi, 20\pi$ and $30\pi$, respectively.
(\textbf{b}) Describes $T_{eff}$ (in units of $mK$) versus $\Delta/\omega_m$ at $\omega/\omega_m=0.1$, 
$G_m/\omega_m=$ and $G_a/\omega_m=5$. Similarly, the black corresponds to $\alpha/\omega_m=0\pi$ while shaded curves from darkest to lightest represent 
$\alpha/\omega_m=60\pi, 
90\pi, 120\pi, 150\pi$ and $170\pi$, respectively. 
(\textbf{c}) and (\textbf{d}) show $T_{eff}$, as a fucntion of $\alpha/\omega_m$ and $U/\omega_m$, at $\Omega_z/\omega_m=0$ and 
$80$, respectively. Here, $G_m/\omega_m=18$, $G_a/\omega_m=16$ and $\alpha/\omega_m=30\pi$.}
\label{fig6}
\end{figure}
However, at $\alpha/\omega_m\neq0\pi$, $S_\downarrow(\omega,\Delta)$ behave 
differently because of more interaction with quantum noise effects, as shown in Fig.\ref{fig2}(e) and \ref{fig2}(f). 
Now the height of semi-circular structure appearing in $S_\downarrow(\omega,\Delta)$ is being increased with increase in SO-coupling. 
The SO-coupling induced Zeeman field effect generates the energy gap between dressed states by increasing and decreasing the ground state energies of pseudo spin-$\downarrow$ and  pseudo spin-$\uparrow$ states, respectively. Therefore, by increasing SO-coupling, spin-$\downarrow$ state will interact with more noise effects and get heated because of having more ground state energy as compared to spin-$\uparrow$ state. 
However, it can be controlled by varying system parameters and the radius of semi-circular structure still appears to be decreasing with SO-coupling due to cavity mediated self-confinement via back-action. Further, in presence of SO-coupling, atom-atom interactions will effect similarly the low-temperature dynamics of atomic mode, as explained in Appendix \ref{ap5}. 

\section{MECHANICAL MIRROR COOLING}
The effective temperature of mechanical 
mode ($T_{eff}$) is calculated by formula $T_{eff}=\langle E_m\rangle/k_B$, where 
$\langle E_m\rangle=m\omega^2_m\langle\delta \hat{q}^2\rangle/2+\langle\delta \hat{p}^2\rangle/2m=m\omega^2_m(n_{eff}+1/2)$, corresponds 
to the mean energy which is experimentally measured by calculating area underneath DNS of mechanical mirror $S_{m}(\omega,\Delta)=\frac{1}{4\pi}\int e^{-i(\omega+\omega^{\prime})t}\langle \delta\hat{q}(\omega)\delta\hat{q}(\omega^{\prime})+\delta \hat{q}(\omega^{\prime})\delta \hat{q}(\omega)\rangle d\omega^{\prime}$, where $\delta\hat{q}$ is dimensionless position quadrature of mechanical mirror defined as, $\delta\hat{q}=\frac{1}{\sqrt{2}}(\hat{b}+\hat{b}^{\dag})$ (for detail see Appendix \ref{ap2} and \ref{ap3}). $n_{eff}$ is the effective phonon number which 
should be less than one in order to achieve ground state cooling. The position and momentum variances are related to DNS, 
$\langle\delta \hat{\mathcal{R}}^2\rangle=\frac{1}{2\pi}\int S_\mathcal{R}(\omega,\Delta)d\omega$, where $\mathcal{R}$ is a generic operator representing position $\delta\hat{q}$ and momentum $\delta\hat{p}$ quadrature of mechanical mirror. Here the DNS of mechanical mirror in momentum space is defined as $S_{m(p)}(\omega,\Delta)=m^2\omega^2_mS_m(\omega,\Delta)$, where $m$ is the effective mass of mechanical mirror.

The cooling mechanism for mechanical mirror, which can simply be explained by thermodynamic arguments, only occurs when the intra-cavity optical sideband is centered at $\omega_m$ which is in fact a resolved-sideband regime. Therefore, the BEC should also oscillate at optical sideband frequency in order to absorb excitation energies of the mirror from cavity mode otherwise mirror temperature will be unaffected. The implication of SO-coupling splits atomic mode into dressed spin states -- acting like two atomic mirrors equally coupled to the cavity -- which will modify atomic back-action inside the cavity. 
This phenomenon enables us to transfer more excitation energies in the form of phonons from mechanical mirror to atomic degree of freedom. Fig.\ref{fig6}(a) illustrates such 
effects where the suppression of mechanical mirror DNS $S_m(\omega,\Delta)$ can be seen with the increase in SO-coupling. 
The SO-coupling suppresses the heating effects induced by the Brownian motion of the mirror and 
enhances back-action cooling of oscillating mirror. Fig.\ref{fig6}(a) demonstrates mirror DNS at system detuning $\Delta/\omega_m=1.8$. If we 
change system detuning, it will modify mirror DNS by increasing or decreasing its strength but the effects of SO-coupling on mirror DNS will remain the same, as discussed in Appendix \ref{ap6}.
Further, the SO-coupling reduces $T_{eff}$ over a wide range of detuning because of 
energy transformation via modified back-action, 
as shown in Fig.\ref{fig6}(b), where optimal temperature is decreasing with the increase in SO-coupling. 
This implies, like atomic-field coupling and atom-atom interactions (as discussed in Appendix \ref{ap7}) 
\cite{Sonam2013,wilson2015,Wilson2007,Wollman2015,nigues2015,liu2013,Elste2009}, SO-coupling significantly alters 
ground state properties of mechanical mirror. Thus, SO-coupling provides another handle to achieve and sustain 
ground state cooling which is even beyond the previous back-action cooling mechanism. 

To further analyze the influence of SO-coupled dressed states on 
mechanical mirror, we plot $T_{eff}$, as a function of 
$U/\omega_m$ and $\alpha/\omega_m$, in absence (Fig.\ref{fig6}(c)) and in presence (Fig.\ref{fig6}(d)) of 
Raman coupling $\Omega_z/\omega_m$. At $\Omega_z/\omega_m=0$, the maximum value of $T_{eff}$ appears to be 
approximately centered at $U/\omega_m\approx6$ and remains saturated with increase in $\alpha/\omega_m$. One can 
state that the maximum value of $T_{eff}$ shows a kind of localized behavior with SO-coupling which is similar to the results presented in reference \cite{Sonam2013}. 
On the other hand, in presence of Raman coupling, $T_{eff}$ shows squeezed and exponential behavior 
with $\alpha/\omega_m$, as illustrated in Fig.\ref{fig6}(d). The SO-coupling in presence of strong Raman coupling, which transforms atomic dispersion spectrum into single minima, absorbs more mirror excitation energies and manipulates atom-atom interactions effects on mirror temperature by modifying back-action. Intuitively saying, the higher values of Raman coupling change quantum phase of trapped atoms causing the alteration in their many-body interactions as well as in SO-coupling effects. 
Therefore, the suppressed and nonlinear behavior of $T_{eff}$ is caused by the emergence of band-gap induced quantum phase transitions of BEC and can be further 
enhanced by increasing $\Omega_z/\omega_m$ providing control over temperature of mechanical mirror \cite{Sonam2013,wilson2015,Wilson2007}. 

\begin{figure}[tp]
	\includegraphics[width=8.7cm]{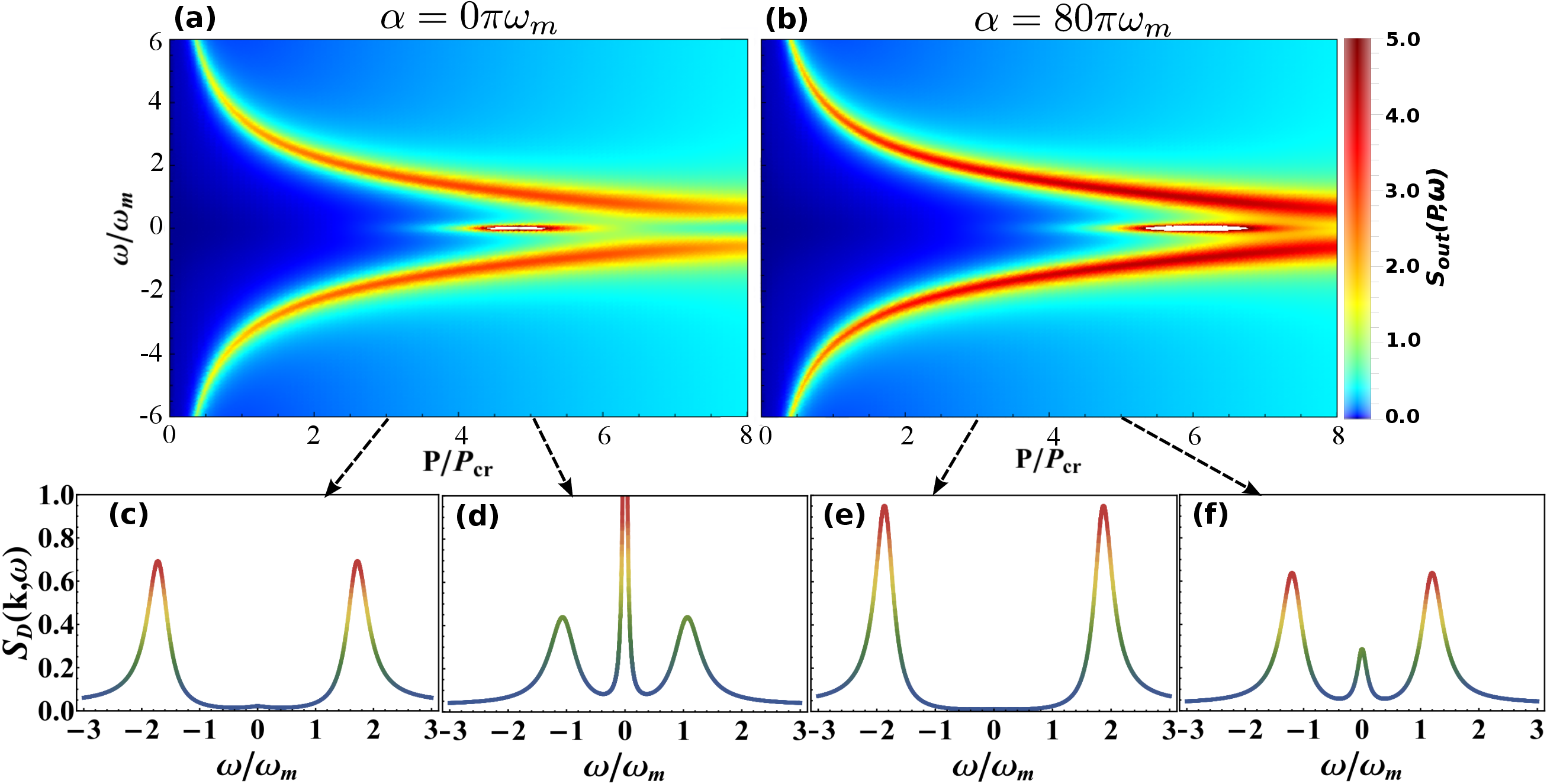}
	\caption{(Color online) (\textbf{a}) and (\textbf{b}) DNS of optical field leaking-out of cavity $S_{out}(P,\omega)$ (in units of $W/Hz$), versus 
		$P/P_{cr}$ and $\omega/\omega_m$, for $\alpha/\omega_m=0\pi$ and $80\pi$, respectively. 
		(\textbf{c-f}) show dynamic structure factor (in units of $1/Hz$) corresponding to the out-going optical mode DNS versus $\omega/\omega_m$ for different values 
		of $P/P_{cr}$. SO-coupling $\alpha/\omega_m=0\pi$ ($80\pi$) for (\textbf{c}) and (\textbf{d}). While the strength of SO-coupling is $\alpha/\omega_m=80\pi$ for (\textbf{e}) and (\textbf{f}). 
		Here $G_m/\omega_m=1.5$, $G_a/\omega_m=0.9$, $U/\omega_m=5.5$ and $\delta/\omega_m=0$.}
	\label{fig7}
\end{figure}
\section{DYNAMIC STRUCTURE FACTOR}
We analyze dynamic structure factor by computing Fourier domain auto-correlations of light leaking-out 
of the cavity. The resultant dynamic structure factor is given as \cite{Landig2015}, 
$S_{D}(k,\omega)=\frac{4(\kappa^2+\Delta^2)}{N\eta^2}(\frac{1}{2\pi}S_{out}(P,\omega)+n_s^2\delta(\omega)),$ 
where $n_s$ is the steady-state photon number and $S_{out}(P,\omega)$ is DNS of out-going optical 
mode (see Appendix \ref{ap4}). The frequency $\omega$ is referred to the shifted frequency of input 
field after interacting with the system which causes inelastic photon scattering. 

In absence of SO-coupling, $S_{out}(P,\omega)$ contains two sidebands 
at $\omega<0\omega_m$ and $\omega>0\omega_m$ 
caused by the incoherent creation and annihilation of quasi-particles \cite{Landig2015}, respectively, 
see Fig.\ref{fig7}(a). 
If we increase the input power, both the sidebands tend to move towards $\omega=0\omega_m$ because of quantum fluctuations which 
decrease the spectral densities of quasi-particles. 
Intuitively, it is referred to 
the scattering of intra-cavity optical mode at Bragg planes in the density-modulated cloud \cite{Birkl1995,miyake2011}. 
Both the sidebands seem to get mixed with each other due to the presence of another secondary structure approximately at $P\approx6P_{cr}$. 
The secondary structure, which is centered at $\omega=0\omega_m$, is caused by associated quantum 
noises \cite{Landig2015}.
In the presence of SO-coupling, secondary structure is shifted to 
$P\approx7P_{cr}$ due to the modifications in the inelastic scattering 
of cavity mode by atomic spin-phase transitions \cite{Punk2014}, 
as shown in Fig.\ref{fig7}(b). 
The spectral densities of both sidebands as well as secondary structure are also increased due to the addition of quasi-particles excited by the 
SO-coupling. 

\begin{figure}[tp]
	\includegraphics[width=8.5cm]{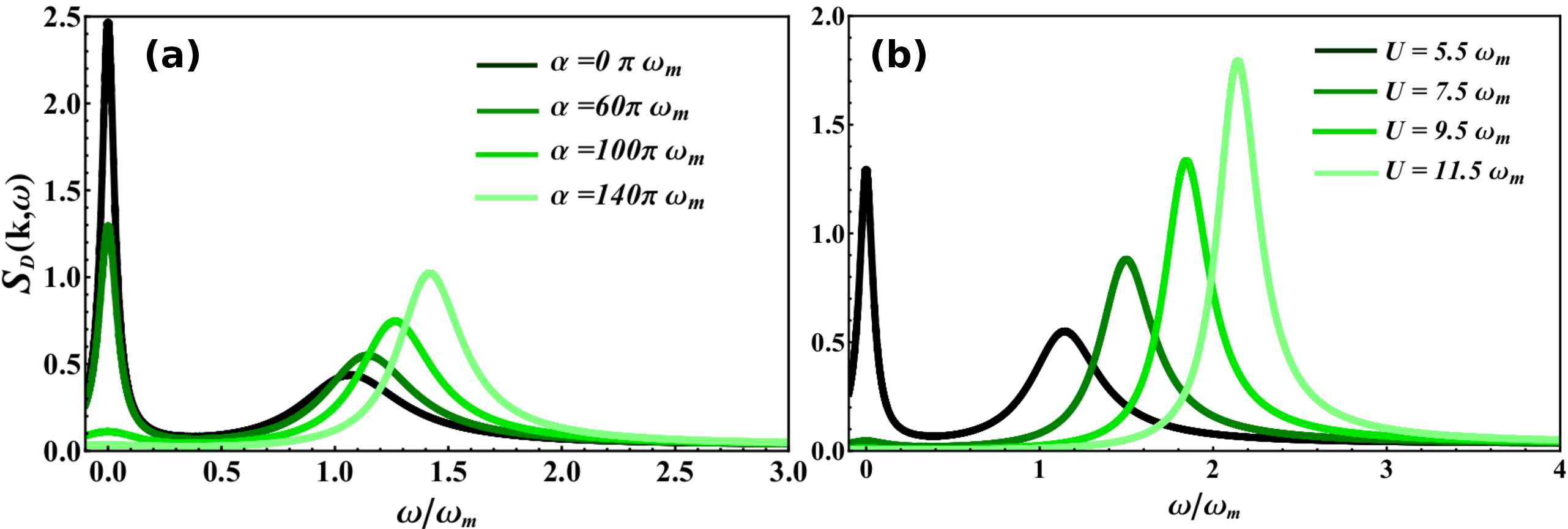}
	\caption{(Color online) The dynamic structure factor $S_{D}(k,\omega)$ (in units of $1/Hz$) as a function of $\omega/\omega_m$ for 
		different strengths of $\alpha/\omega_m$ at $U=5.5\omega_m$ (\textbf{a}) 
		and atom-atom interactions $U/\omega_m$ at $\alpha=10\pi\omega_m$ (\textbf{b}). The input field power ratio is fixed to $P/P_{cr}=4$ and the remaining coupling strengths 
		are same as in Fig.\ref{fig7}. In (\textbf{a}), the green shaded curves from darkest to lightest correspond to the SO-coupling $\alpha/\omega_m=0\pi, 
		60\pi, 100\pi$ and $140\pi$, respectively. 
		Similarly, in (\textbf{b}), green curves (from dark shade to light shade) carry the influence of atom-atom interactions with strengths $U/\omega_m=5.5, 
		7.5, 9.5$ and $11.5$, respectively.}
	\label{fig9}
\end{figure}
Dynamic structure factor $S_{D}(k,\omega)$ at power ratio 
$P/P_{cr}=3$ shows two sidebands at $\omega\approx-2\omega_m$ and $\omega\approx2\omega_m$ corresponding 
to creation and annihilation of quasi-particles, 
respectively, as shown in Fig.\ref{fig7}(c). Another, comparatively small, fluctuating structure can be 
seen at $\omega=0\omega_m$ induced by the quantum-noise effects which 
verifies the experimental finding of dynamic structure factor in reference \cite{Landig2015}. If we increase the 
input field power, the dynamic structure factor will be suppressed by the increase in system fluctuations 
and the sidebands will move towards $\omega=0\omega_m$, as shown in Fig.\ref{fig7}(d). 
However, the strength of structure appearing at $\omega=0\omega_m$ is increased because of quantum noise effects. 
Interestingly, these effects can be suppressed by SO-coupling because of enhanced intra-cavity atomic back-action, 
see Fig.\ref{fig7}(e) and \ref{fig7}(f), which leads to the enhancement of sideband spectrum. 

In order to further understand influence of SO-coupling as well as cavity mediated long-range atom-atom 
interactions, we plot $S_{D}(k,\omega)$ for multiple values of 
SO-coupling and atom-atom interactions. Fig.\ref{fig9}(a) and \ref{fig9}(b) carry $S_{D}(k,\omega)$ as a function of normalized 
frequency for different strengths of $\alpha/\omega_m$ and $U/\omega_m$, respectively, at input field power $P/P_{cr}=5$. 
It can be clearly seen that by increasing the SO-coupling and atom-atom interactions, the sidebands are enhanced and shifted away from 
$\omega=0\omega_m$ due to the addition of quasi-particles. However, quantum noise fluctuations, appearing at $\omega\approx0$, are 
now being suppressed by increasing $\alpha/\omega_m$ and $U/\omega_m$ causing enhancement in optomechancial applications. 
Here, it should be noted that the strength of atom-atom interactions defines inter-species as well as intra-species interactions and the frequency of the atomic mode is directly proportional to the strength of interactions $\sqrt{U}$ \cite{Sonam2013}. Therefore, like SO-coupling, atom-atom interactions have significant influence on the intra-cavity atomic back-action which leads to the enhancement in inelastic scattering of the cavity mode.  
Thus, the inclusion of SO-coupling purifies dynamic structure factor by squeezing quantum noises.\\

\section{CONCLUSION}
We demonstrate SO-coupling induced back-action cooling in cavity-optomechanics with SO-coupled BEC. 
The SO-coupling modifies dynamical back-action which enhances low-temperature profile of atomic mode by squeezing 
associated noises. It has been shown that the existence of SO-coupling leads to cool vibrating end mirror to its quantum mechanical 
ground state. Further, by computing dynamic structure factor, we have shown that the SO-coupling 
enables us to manage and implement noiseless quasi-particles. 
Likewise, mechanical mirror gives rise to the eigenenergy spectrum of atomic states providing 
control over quantum phase transitions. We chose a particular set of parameters and 
procedures very close to the present experimental ventures which makes our study experimentally feasible. Our findings constitute a 
significant step towards the utilization of SO-coupled BEC-optomechanics in the field of 
quantum optics and quantum information.

\section*{ACKNOWLEDGMENT}
This work was supported by the NKRDP under grants Nos. 2016YFA0301500, NSFC under grants Nos. 11434015, 61227902, 61378017, KZ201610005011, SKLQOQOD under grants No. KF201403, SPRPCAS under grants No. XDB01020300, XDB21030300. 
We also acknowledge the financial support from CAS-TWAS president's fellowship programme (2014).

\appendix
\section{Atomic eigenenergies calculation}\label{ap1}
Here we provide some details about energy dispersion calculation of atomic states. By adopting a mean-field approximation, we consider the intra-cavity field in steady-state and replace the intra-cavity field 
operator by its expectation value $\hat{c}\rightarrow\langle c\rangle\equiv c_s$. To calculate energy dispersion $E_N$ of atomic modes, 
we define $E_N$ as the solution of nonlinear quantum Langevin equations and replace the time derivative $id/dt$ with eigenenergy 
$E_N$. After performing some mathematics and applying Pauli matrices, the coupled Langevin equations will take the form \cite{Dong2014,Lin2009,Dalibard2011}, 
\begin{widetext} 
\begin{eqnarray}
n_s&=&c_{s}^\dag c_{s}=\frac{\eta}{\kappa^2 +(\tilde{\Delta}-\frac{g_m}{\sqrt{2}}(\hat{b}^\dag+\hat{b})  +g_{a}(\hat{\varphi}^\dag_{\uparrow}\hat{\varphi}_{\uparrow}+
	\hat{\varphi}^\dag_{\downarrow}\hat{\varphi}_{\downarrow}))^2},\label{1}\\
\hat{b}&=&\frac{g_m c_s^{\dag}c_s}{\sqrt{2}(E_N+i\omega_{m}+\gamma_m)},\hat{b}^\dag=\frac{g_m c_s^{\dag}c_s}{\sqrt{2}(E_N-i\omega_{m}+\gamma_m)},\\
E_N\binom{\hat{\varphi}_\uparrow}{\hat{\varphi}_\downarrow}&=&\begin{pmatrix}
\frac{\hbar\pmb{k}^2}{2m}+\frac{\Omega_z}{2}+g_ac_s^{\dag}c_s+\frac{1}{2}UN-\gamma_a&-i(\alpha\pmb{k_x}+\frac{\delta}{2})+\frac{1}{2}U(\varepsilon-1)\hat{\varphi}_\downarrow^\dag\hat{\varphi}_\uparrow\\ 
i(\alpha\pmb{k_x}+\frac{\delta}{2})+\frac{1}{2}U(\varepsilon-1)\hat{\varphi}_\uparrow^\dag\hat{\varphi}_\downarrow&
\frac{\hbar\pmb{k}^2}{2m}-\frac{\Omega_z}{2}+g_ac_s^{\dag}c_s+\frac{1}{2}UN-\gamma_a
\end{pmatrix}\binom{\hat{\varphi}_\uparrow}{\hat{\varphi}_\downarrow},\label{10}
\end{eqnarray}
\end{widetext}
where $n_s$ is the steady-state photon number inside cavity. For simplicity, we have ignored quantum noises associated with the system 
while calculating eigenenergies of atomic mode. Under number conversation condition $|\hat{\varphi}_\uparrow|^2+|\hat{\varphi}_\downarrow|^2=1$, we substitute steady-state mechanical mirror operators into equ.\ref{1} and simplify in term of $n_s$ as,  
\begin{eqnarray}
n_s^3+2L_1L_2n_s^2+Kn_s&=&\eta^2\label{6}
\end{eqnarray}
where,
\begin{eqnarray}
L_1&=&\Delta_c+g_a,\\
L_2&=&\frac{g_m^2(E_N+\gamma_m)}{(E-n+\gamma_m)^2+\omega_m^2},\\
K&=&\kappa^2+L_1^2.
\end{eqnarray}
Now, by assuming eigenenergies of moving-end mirror and atomic mode independent by keeping mechanical mirror in steady-state, we rewrite equ.\ref{10}, for $\hat{\varphi}_\uparrow$ and $\hat{\varphi}_\downarrow$, as,
\begin{small}
\begin{eqnarray}
&&E_N\binom{\hat{\varphi}_\uparrow}{\hat{\varphi}_\downarrow}=\nonumber\\ 
&&\begin{pmatrix}
h_1& w+\frac{1}{2}U(\varepsilon-1)\hat{\varphi}_\downarrow^\dag\hat{\varphi}_\uparrow\\ 
w^*+\frac{1}{2}U(\varepsilon-1)\hat{\varphi}_\downarrow^\dag\hat{\varphi}_\uparrow&
h_2
\end{pmatrix}\binom{\hat{\varphi}_\uparrow}{\hat{\varphi}_\downarrow},\label{1111}
\end{eqnarray}
\end{small}
where, 
\begin{eqnarray}
h_{1,2}&=&\frac{\hbar\pmb{k_x}^2}{2m}\pm\frac{\Omega_z}{2}+g_a n_s+\frac{1}{2}UN-\gamma_a,\\
w&=&\alpha\pmb{k_x}+\frac{\delta}{2}.
\end{eqnarray}
By dividing first line of equ.\ref{1111} with the conjugate of the second line, we obtain, 
\begin{eqnarray}
\frac{E_N-h_1}{E_N-h_2}|\hat{\varphi}_\uparrow|^2&=&|\hat{\varphi}_\downarrow|^2.
\end{eqnarray}
Further, by denoting $s=\frac{E_N-h_1}{E_N-h_2}$ and using number conversation condition, we calculate $|\hat{\varphi}_\uparrow|$ and $|\hat{\varphi}_\downarrow|$ as,
\begin{eqnarray}
|\hat{\varphi}_\uparrow|^2&=&\frac{1}{s+1},\\
|\hat{\varphi}_\downarrow|^2&=&\frac{s}{s+1},
\end{eqnarray}
and by substituting these values in the first line of  equ.\ref{1111}, we obtain,
\begin{eqnarray}
(E_N-h_1)^2&&-\frac{U(\epsilon-1)(E_N-h_2)s}{s+1}+\frac{U^2(\epsilon-1)^2s^2}{4(s+1)^2}\nonumber\\
&& =sw^2.\label{en}
\end{eqnarray}
Finally, by numerically finding roots of steady-state photon number $n_s$ from equ.\ref{6} and substituting them into equ.\ref{en}, 
we plot the roots of eigenenergies versus 
quasi-momentum $\pmb{k_x}$, as shown in Fig.\ref{fig1}. (Note that we consider $\pmb{k_y}=\pmb{k_z}=0$ because SO-coupling is occurring only in the direction of 
$\hat{x}$-axis.)

\section{Langevin equations and frequency domain solutions}\label{ap2}
The coupled Langevin equations of the system contain nonlinear terms in the form of coupling among different degrees of 
freedom and noises associated with the system. By considering intense external pump field, these equations can be 
linearized with the help of quantum fluctuations as, $\mathcal{\hat{O}}(t)=\mathcal{O}_{s}+\mathcal{\delta O}(t)$, 
where $\mathcal{O}$ can be any operator of the system, $\mathcal{O}_{s}$ represents steady-state value and $\mathcal{\delta O}(t)$ 
is the first order quantum fluctuation. During these calculation for simplicity, we assume that both the atomic states, spin-$\uparrow$ and spin-$\downarrow$, 
have equal amount of particles, i.e $\hat{\varphi_\uparrow}^\dag\hat{\varphi_\uparrow}=\hat{\varphi_\downarrow}^\dag\hat{\varphi_\downarrow}=N/2$. 
Furthermore, we define system quadratures in the form of dimensionless position and momentum 
quadratures as, $\hat{q}_O=\frac{1}{\sqrt{2}}(\hat{O}+\hat{O}^{\dag})$ and $\hat{p}_O=\frac{i}{\sqrt{2}}(\hat{O}-\hat{O}^{\dag})$, 
respectively, ($O$ is generic operator) having commutation relation $[\hat{q}_O,\hat{p}_O]=i$ which 
reveals the value of scaled Planck's constant $\hbar=1$. 
Now the linearized Langevin equation are defined in form of $\mathcal{\dot{X}=KX+F}$, where vector 
$\mathcal{X}=[\delta q_c(t),\delta p_c(t),\delta q(t),\delta p(t),\delta q_{\uparrow}(t),\delta p_{\uparrow}(t),\delta q_{\downarrow}(t),\delta p_{\downarrow}(t)]^\tau$ 
contains position and momentum quadratures of the system (here $p$ and $q$ with $\uparrow$ and $\downarrow$ indicate atomic states, with $c$ indicates cavity mode and without anything indicate mechanical mirror's momentum and position quadrature) and vector $\mathcal{F}=[\sqrt{2\kappa}q^{in}_c,\sqrt{2\kappa}p^{in}_c,0,2\sqrt{\gamma_m}f_m,0,2\sqrt{\gamma_a}f_a,0,2\sqrt{\gamma_a}f_a]^\tau$ defines noises associated with the system. The matrix $\mathcal{K}$ contains dynamical parameters associated with the 
system, 
\begin{widetext} 
\begin{center}
	$\mathcal{K}=
	\begin{pmatrix}
	-\kappa & \Delta & 0 & 0 & 0 & 0 & 0 & 0 \\
	\Delta & -\kappa & -G_m & 0 & G_a & 0 & G_a & 0 \\
	-2G_m & 0 & -\gamma_m & \omega_m & 0 & 0 & 0 & 0 \\
	0 & 0 & -\omega_m & -\gamma_m & 0 & 0 & 0 & 0 \\
	2G_a & 0 & 0 & 0 & M & \frac{\Omega_z}{2}& (\alpha-\frac{\delta}{2}) & 0 \\
	0 & 0 & 0 & 0 &  \frac{\Omega_z}{2} & M & 0 & -(\alpha-\frac{\delta}{2}) \\
	2G_a & 0 & 0 & 0 & (-\alpha+\frac{\delta}{2})& 0 & M &  -\frac{\Omega_z}{2} \\
	0 & 0 & 0 & 0 & 0 & -(-\alpha+\frac{\delta}{2}) & -\frac{\Omega_z}{2} & M
	\end{pmatrix}$,
\end{center}
\end{widetext}
where $M=\frac{\Omega}{2}+v+UN(1-\varepsilon)-\gamma_a$, $v=g_an_s$ and $\Omega=\hbar\pmb{k}^2/m_a$ is the recoil frequency of 
atomic states. $\alpha=\tilde{\alpha}\pmb{k_x}$ is the effective strength of SO-coupling. 
The evolution of the system can be analyzed by matrix $\mathcal{K}$ which contains multiple crucial parameters 
such as effective detuning $\Delta = \tilde{\Delta}-g_m q_s +g_{a}N$, where $q_s$ is steady-state quadratures of mechanical mirror, 
and modified coupling of intra-cavity optical mode with mechanical mirror $G_{m}=\sqrt{2}g_m|c_{s}|$ and atomic modes $G_{a}=\sqrt{2}g_{a}|c_{s}|$, 
tuned by the mean intra-cavity field with amplitude $c_{s}=\frac{\eta}{\kappa +i\Delta}$.
The particular interlaced nature of these steady-state parameters provides an efficient 
opportunity to understand nonlinear and bistable dynamics of the system.

To make the system accurate and useful, we have to 
ensure stability of the system and for this purpose, we perform stability analysis of the system. 
The system can only be stable if the roots of the characteristic polynomial of 
matrix $\mathcal{K}$ lie in the left half of the complex plane. For this purpose, we apply Routh-Hurwitz Stability Criterion 
\cite{peter2} on matrix $\mathcal{K}$ and numerically develop stability conditions for the system. 
These stability conditions are given as, 
$M>\kappa+\gamma_m$, $(\alpha-\delta/2)^2+M^2>\kappa^2+\Delta^2-\omega_m^2-\Omega_z^2$,  
$\omega_m>\Delta>\kappa>\gamma_m>0$ and $\Delta G_a^2+\Delta G_m^2>M(\kappa^2-\Omega_z^2)$.
We strictly follow these conditions while performing all numerical calculations in the manuscript. 

Furthermore, we take Fourier 
transform of linearized Langevin equations to preform 
frequency domain analysis and solve them for position and momentum quadratures of intra-cavity field, 
\begin{eqnarray}
\delta q_c(\omega)&=&\frac{1}{L(\omega)}\bigg(\sqrt{2\kappa}[\Delta\delta p^{in}_c+(\kappa+i\omega)\delta q^{in}_c]\nonumber\\
&+&\Delta [G_a\delta q_\uparrow(\omega)+G_a\delta q_\downarrow(\omega) -G_m\delta q(\omega)]\bigg),\\
\delta p_c(\omega)&=&\frac{1}{L(\omega)}\bigg(\sqrt{2\kappa}[\Delta\delta q^{in}_c+(\kappa+i\omega)\delta p^{in}_c]+(\kappa+i\omega)\nonumber\\
&&[G_a\delta q_\uparrow(\omega)+G_a\delta q_\downarrow(\omega) -G_m\delta q(\omega)]\bigg),
\end{eqnarray}
respectively, position quadrature of atomic modes,
\begin{eqnarray}
\delta q_{\uparrow,\downarrow}(\omega)&=&\frac{1}{X(\omega)}\bigg((B_{\uparrow,\downarrow}
(\omega)+A_{\downarrow,\uparrow}(\omega))C(\omega)[\Delta\delta p^{in}_c+(\kappa\nonumber\\
&&+i\omega)\delta q^{in}_c]+L_{1,3}(\omega)f_m
+L_{2,4}(\omega)f_a\bigg),
\end{eqnarray}
and finally for the position quadrature of mechanical mirror,
\begin{eqnarray}
\delta q(\omega)&=&\frac{1}{X_m(\omega)}\bigg(A_m(\omega)[\Delta\delta p^{in}_c
+(\kappa+i\omega)\delta q^{in}_c]\nonumber\\
&&+B_{m}(\omega)f_m
+C_{m}(\omega)f_a\bigg).
\end{eqnarray}
The parameter $L(\omega)=(\kappa+i\omega)^2-\Delta^2$ contains effective detuning of the system, 
$W(\omega)=\gamma_a+i\omega-\Omega/2-v-UN(1-\varepsilon)$,$K(\omega)=W^2(\omega)+(\alpha^2-\delta/2)^2$ describes atom-atom interactions and 
$S_m(\omega)=(\gamma_m+i\omega)^2L(\omega)-L(\omega)\omega^2_m+2G^2_m\Delta(\gamma_m+i\omega)$ is related to mirror coupling with intra-cavity field. 
$A_{\uparrow,\downarrow}(\omega)=4W(\omega)K(\omega)L(\omega)S_m(\omega)\pm\Omega^2_zL(\omega)S_m(\omega)-8G_a^2\Delta K(\omega)S_m(\omega)
+16G_a^2\Delta^2G_m^2(\gamma_m+i\omega)K(\omega)$ and $B_{\uparrow,\downarrow}(\omega)=\pm\Omega^2_zL(\omega)S_m(\omega)
+4(\pm\alpha\mp\delta/2)K(\omega)L(\omega)S_m(\omega)+8G_a^2\Delta K(\omega)S_m(\omega)
-16G_a^2\Delta^2G_m^2(\gamma_m+i\omega)K(\omega)$ describes the behavior of atomic mode and its association with moving-end mirror of 
the system. $B_m(\omega)=2G_m\sqrt{(2\kappa)}(\gamma_m+i\omega)X(\omega)+=2G_m\Delta G_a(A_{\uparrow}(\omega)+A_{\downarrow}(\omega)
+B_{\uparrow}(\omega)+B_{\downarrow}(\omega))$ represents mechanical mirror behavior and its coupling with atomic modes. 
Further, $C(\omega)=8G_a^2\sqrt{(2\kappa)}K(\omega)S_m(\omega)+16G_a^2\sqrt{2\kappa}G_m^2(\gamma_m+i\omega)K(\omega)$, 
$L_{1,3}(\omega)=(B_{\uparrow,\downarrow}(\omega)+A_{\downarrow,\uparrow}(\omega))8G_a^2\sqrt{\gamma_m}\Delta K(\omega)L(\omega)(\gamma_m+i\omega)$, 
$L_{2,4}(\omega)=(B_{\uparrow,\downarrow}(\omega)+A_{\downarrow,\uparrow}(\omega))8G_a^2\sqrt{\gamma_a}K(\omega)S_m(\omega)$, 
$B_m(\omega)=2G_m\Delta G_a(L_1(\omega)+L_3(\omega))+2\sqrt{\gamma_m}L(\omega)X(\omega)$ and $C_m(\omega)=2G_m\Delta G_a(L_2(\omega)+L_4(\omega))$.
The term $X(\omega)=A_{\uparrow}(\omega)A_{\downarrow}(\omega)+B_{\uparrow}(\omega)B_{\downarrow}(\omega)$ 
represents modified susceptibility atomic states and $X_m(\omega)=X(\omega)S_m(\omega)$ corresponds to the modified susceptibility of mechanical mirror.

\section{Density-noise spectrum (DNS)}\label{ap3}
By using frequency domain solutions given above and standard formalism for auto-correlation, 
$S_{\mathcal{O}}=\frac{1}{2\pi}\int e^{-i(\omega-\acute{\omega})}\langle \mathcal{O}(\omega)\mathcal{O}(\acute{\omega})\rangle d\acute{\omega}$, as discussed in main text, 
where $\mathcal{O}(\omega)$ is the generic operator, the DNS for pseudo spin-$\uparrow$ and 
spin-$\downarrow$ atomic states will be read as,
\begin{eqnarray}
S_{\uparrow,\downarrow}(\omega,\Delta)&=&\frac{1}{|X(\omega)|^2}\bigg(2\pi|C(\omega)|^2(|B_{\uparrow,\downarrow}
(\omega)|^2+\nonumber\\
&&|A_{\downarrow,\uparrow}(\omega)|^2)[\Delta^2+\kappa^2+\omega^2]+2\pi L_{2,4}(\omega)\nonumber\\
&&+L_{1,3}(\omega)\frac{\gamma_m\omega}{\omega_m}[1+Coth(\frac{\hbar\omega}{2k_BT})]
\bigg).
\end{eqnarray}
Similarly, we can write DNS equation for mechanical mirror of the system as, 
\begin{eqnarray}
S_m(\omega,\Delta)&=&\frac{1}{| X_m(\omega)|^2}\Bigg(|A_m(\omega)|^2(\Delta^2+\kappa^2+\omega^2)\nonumber\\
&&+2\pi B_{m}(\omega)+C_{m}(\omega)\frac{\gamma_m\omega}{\omega_m}[1\nonumber\\
&&+Coth(\frac{\hbar\omega}{2k_BT})]\Bigg).
\end{eqnarray}

\section{Spectral density of out-going optical field}\label{ap4}
In order to calculate output optical field of the system, we use 
input-output field relation, $\delta q_c^{out}=\sqrt{2k}\delta q_c-\delta q_c^{in}$ and $\delta p_c^{out}=\sqrt{2k}\delta p_c-\delta p_c^{in}$, 
where $p_{in},q_{in}$ and $p_{out},q_{out}$ represent input and output field quadratures, 
respectively. By utilizing above relation and intra-cavity field quadrature, we obtain output field relation as,  
\begin{eqnarray}
\delta q^{out}_c(\omega)&=&\frac{1}{L(\omega)}\Bigg(\big[2\kappa\Delta\delta p^{in}_c+(\kappa^2+\omega^2+\Delta^2)\delta q^{in}_c\big]\nonumber\\
&&+\sqrt{2\kappa}\Delta \big[G_a \delta q_\uparrow(\omega)+G_a \delta q_\downarrow(\omega)\nonumber\\
&& -G_m\delta q(\omega)\big]\Bigg),\\
\delta p^{out}_c(\omega)&=&\frac{1}{L(\omega)}\Bigg(\big[2\kappa\Delta\delta q^{in}_c+(\kappa^2+\omega^2+\Delta^2)\delta p^{in}_c\big]\nonumber\\
&&+\sqrt{2\kappa}(\kappa+i\omega)\big[G_a \delta q_\uparrow(\omega)+G_a\delta q_\downarrow(\omega)\nonumber\\
&& -G_m\delta q(\omega)\big]\Bigg).
\end{eqnarray}
Now, by combining position and momentum quadratures of field, we obtain out-going field operator $c_{out}$ as,
\begin{eqnarray}
\delta c_{out}(\omega)&=&\frac{1}{L(\omega)}\Bigg(\big[2\kappa\Delta\delta c^\dag_{in}+(\kappa^2+\omega^2+\Delta^2)\delta c_{in}\big]\nonumber\\
&&+\sqrt{2\kappa}\Delta \big[G_a\delta q_\uparrow(\omega)+G_a\delta q_\downarrow(\omega)\nonumber\\ 
&&-G_m\delta q(\omega)\big]\Bigg).
\end{eqnarray}
Further, to determine the dependence of out-going optical mode on the external pump field power $P$, we redefine coupling terms as a function of 
$P$, 
\begin{eqnarray}
G_m&=&\sqrt{2}C_Sg_m=\frac{2\omega_c}{L}\sqrt{\frac{P\kappa}{m\omega_m\omega_p(\kappa^2+\Delta^2)^2}},\\
G_a&=&\sqrt{2}C_Sg_a=\frac{2\omega_c}{L}\sqrt{\frac{P\kappa}{m\Omega\omega_p(\kappa^2+\Delta^2)^2}},
\end{eqnarray}
and after this, we calculate Density-noise spectrum (DNS) of out-going optical mode by simply using two frequency auto-correlation formula,
\begin{eqnarray}
S_{out}(P,\omega)&=&\frac{2\pi}{|L(\omega)|^2}\Bigg(\big[\kappa^2+\omega^2+\Delta^2+2\kappa\Delta\big]\nonumber\\
&&+4\kappa\Delta \big[G_a S_\uparrow(\omega,\Delta)+G_a S_\downarrow(\omega,\Delta)\nonumber\\ 
&&-G_m\delta S_m(\omega,\Delta)\big]\Bigg).
\end{eqnarray}

\section{Influence of atom-atom interactions on atomic density-noise spectrum}\label{ap5}
\begin{figure}[h]
	\includegraphics[width=8.5cm]{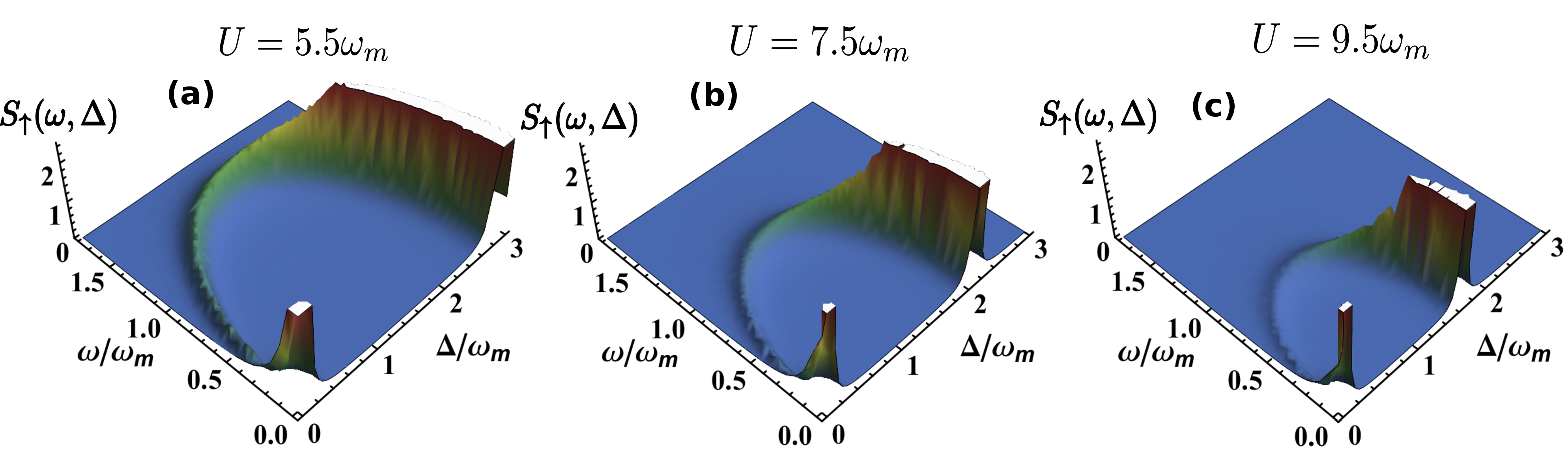}
	\caption{(Color online) The dynamics of $S_\uparrow(\omega,\Delta)$, as a function of detuning $\Delta/\omega_m$ and 
		frequency $\omega/\omega_m$, under the effects of many-body interactions $U/\omega_m$. 
		(\textbf{a}), (\textbf{b}) and (\textbf{c}) demonstrate $S_\uparrow(\omega,\Delta)$ with atom-atom 
		interaction $U=5.5\omega_m$, $7.5\omega_m$ and $9.5\omega_m$, respectively. Here, the strength of SO-coupling is kept constant 
		$\alpha=10\pi\omega_m$. One can observe that the strength of atom-atom interactions influences $S_\uparrow(\omega,\Delta)$ 
		in a similar way as SO-coupling does. By increasing $U/\omega_m$, the area underneath $S_\uparrow(\omega,\Delta)$ is decreased which leads to 
		the cooling of atomic mode \cite{Sonam2013}. The remaining parameters are same as in Fig.\ref{fig1}.}
	\label{sfig1}
\end{figure}
The atom-atom interactions $U/\omega_m$ of 
atomic dressed states show similar influence on $S_\uparrow(\omega,\Delta)$ as the influence of SO-coupling on the atomic dressed states, 
which can be seen in Fig.\ref{sfig1}(a-c), where 
the strength of atom-atom interactions is considered as, $U=5.5\omega_m, 7.5\omega_m, 9.5\omega_m$, respectively. 
The radius as well as height of the atomic DNS decreases with increase in atom-atom interactions $U/\omega_m$ of dressed states. 
(Note: The effects of atom-atom interactions $S_\downarrow(\omega,\Delta)$ are not shown here because they will be like-wise as on 
$S_\uparrow(\omega,\Delta)$.) As atom-atom interactions are the combination of inter-species as well as intra-species interactions and modifies the coupling atomic states with intra-cavity potential, therefore, by increasing interactions, the strength of atomic back-action will be increased leading to more self confinement. 
Thus, the strength of atom-atom interactions can likely be used to control the low-temperature dynamics of 
atomic dressed states as SO-coupling.

\section{Mirror density-noise spectrum under influence of SO-coupling and atom-atom interactions}\label{ap6}
\begin{figure}[h]
	\includegraphics[width=8.5cm]{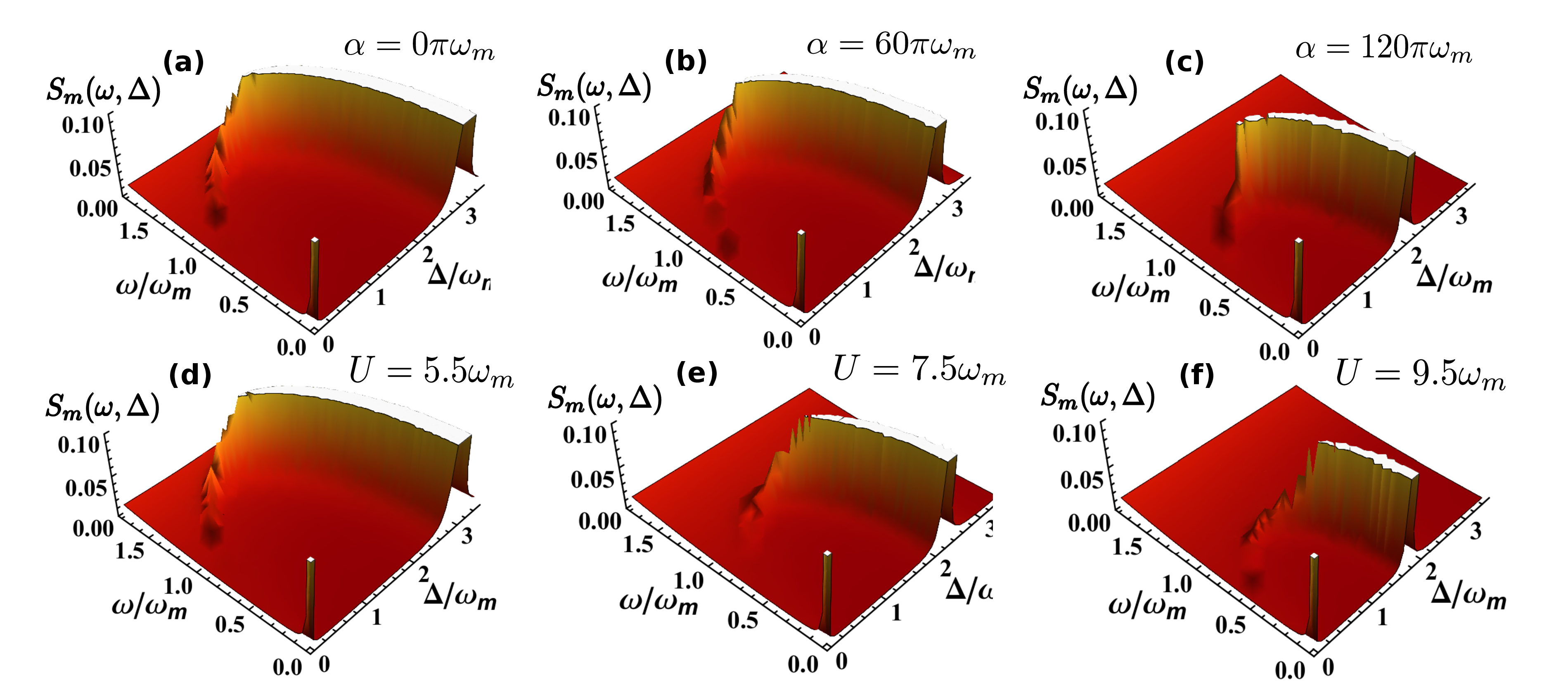}
	\caption{(Color online) Density-noise spectrum (DNS) $S_m(\omega,\Delta)$ (in units of $W/Hz$) for mechanical mirror of the system versus normalized effective detuning $\Delta/\omega_m$ and 
		frequency $\omega/\omega_m$ under the influence of spin-orbit (SO) coupling $\alpha/\omega_m$ of atomic spin-states and normalized atom-atom interactions $U/\omega_m$ 
		among atomic states. (\textbf{a}) Demonstrates $S_m(\omega,\Delta)$ in the absence of SO-coupling $\alpha=0\pi\omega_m$ 
		with atom-atom 
		interactions $U=5.5\omega_m$. The color configuration corresponds to the strength of mechanical mirror DNS ($S_m(\omega,\Delta)$). 
		(\textbf{b}) and (\textbf{c}) illustrate the behavior of $S_m(\omega,\Delta)$ under the influence of $\alpha=60\pi\omega_m$ and $120\pi\omega_m$, respectively. 
		The dynamics of $S_m(\omega,\Delta)$ under the effects of many-body interactions $U/\omega_m$ are illustrated in 
		(\textbf{d-f}). (\textbf{d}), (\textbf{e}) and (\textbf{f}) demonstrate $S_m(\omega,\Delta)$ as a function of 
		$\Delta/\omega_m$ and $\omega/\omega_m$ with atom-atom 
		interaction $U=5.5\omega_m$, $7.5\omega_m$ and $9.5\omega_m$, respectively, while the strength of SO-coupling is kept constant 
		$\alpha=10\pi\omega_m$. The atom-field coupling is considered as $G_a=4.1\omega_m$ while 
		the mirror-field coupling is taken as $G_m=1.5\omega_m$. The remaining parameters, used in numerical calculations, 
		are same as in Fig.\ref{fig1}.}
	\label{sfig7}
\end{figure}
The dynamics of mechanical mirror will also be influenced by the existence of atomic-states as 
atomic dressed states are influenced by the existence of mechanical mirror. 
Fig.\ref{sfig7} demonstrates 
$S_m(\omega,\Delta)$ as a function of $\Delta/\omega_m$ and $\omega/\omega_m$, under the 
influence of SO-coupling and atom-atom interaction. The atom-field coupling is considered as $G_a=4.1\omega_m$ while 
the mirror-field coupling is taken as $G_m=1.5\omega_m$. 
The behavior of $S_m(\omega,\Delta)$ in the absence of SO-coupling is 
shown in Fig.\ref{sfig7}(a), which is similar to the behavior of atomic DNS. 
A semi-circular structure appears with increase in detuning $\Delta/\omega_m$ towards frequency $\omega/\omega_m$. 
The height of $S_m(\omega,\Delta)$ decreases initially and achieves optimal cooling point. 
However, when the system detuning is further increased from $\Delta=1\omega_m$, the $S_m(\omega,\Delta)$ 
shows rapid increase in the height of structure giving rise to the temperature of mechanical mirror. 

The strength of SO-coupling 
induces similar influence as it does on the atomic DNS. The radius of the structure is suppressed by the increased SO-coupling, 
as shown in Fig.\ref{sfig7}(b) and \ref{sfig7}(c), where the strength of SO-coupling is increased to $\alpha=60\omega_m$ and $130\omega_m$, 
respectively. Not only SO-coupling but also the atom-atom interactions of atomic dressed states will show similar effects on mechanical DNS 
as they are inducing in atomic DNS. The increase in atom-atom interactions will also reduce the radius of semi-circular 
structure, as shown in Fig.\ref{sfig7}(d), \ref{sfig7}(e) and \ref{sfig7}(f), where the strength of atom-atom interactions is increased to 
$U=5.5\omega_m, 7.7\omega_m$ and $9.5\omega_m$, respectively. The SO-coupling and atom-atom interactions modify the 
atomic density mode excitation leading to the variation in intra-cavity optical spectrum in the form of modified atomic back-action which will consequently lead to the absorption of more mirror excitations by spin states. 
It can also be considered as the atomic and mechanical states 
are connected with each other through intra-cavity radiation pressure, acting as a spring between these two independent entities, 
therefore, the modifications produced by SO-coupling and atom-atom interaction will show similar influence 
on mechanical mirror as they are producing on atomic dressed states \cite{Sonam2013}.

\section{Influence of atom-field coupling and atom-atom interactions on mechanical mirror temperature}\label{ap7}
\begin{figure}[h]
	\includegraphics[width=8.4cm]{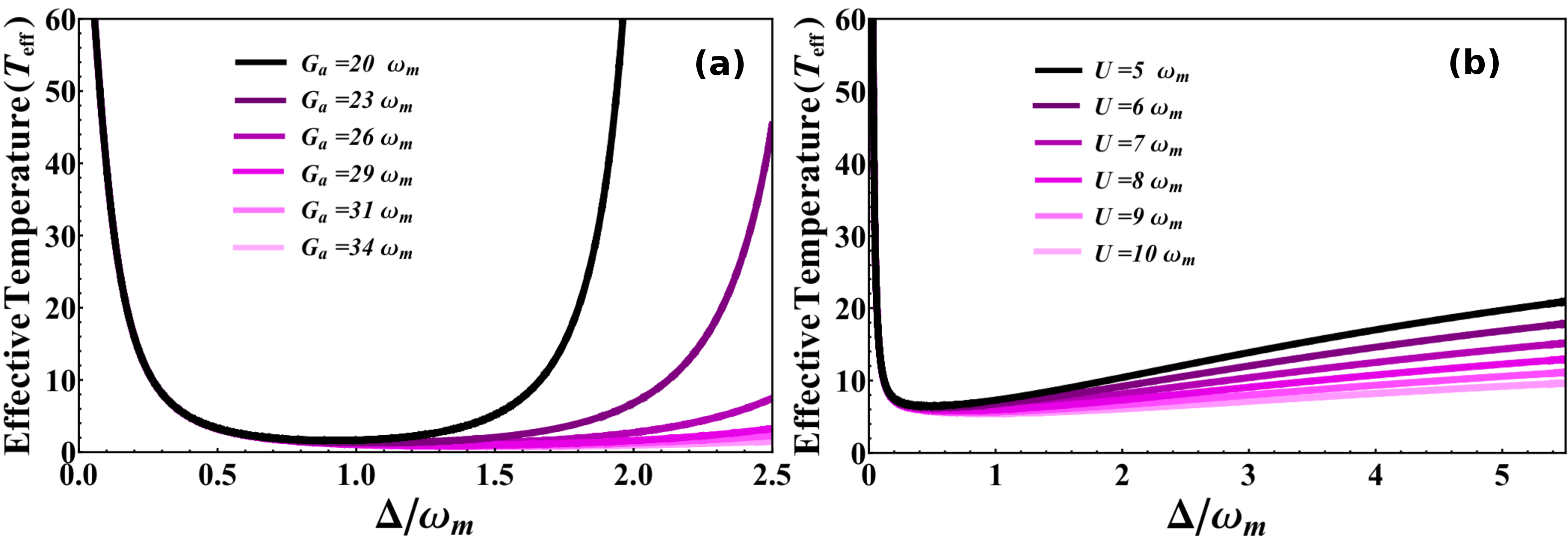}
	\caption{(Color online) (\textbf{a}) Illustrates the effective-temperature $T_{eff}$ of mechanical mirror under the 
		influence of $G_a/\omega_m$. The SO-coupling strength is now considered as $\alpha=100\pi\omega_m$ 
		and many-body interaction is kept as $U=5.5\omega_m$. The black curve represents $G_a=20\omega_m$ while magenta curves from dark shade to light shade are for 
		atom-field coupling $G_a=23\omega_m, 26\omega_m, 29\omega_m, 31\omega_m$ and $34\omega_m$, respectively. Similarly, 
		(\textbf{c}) deals with the behavior effective-temperature $T_{eff}$ of mechanical mirror under the influence of $U/\omega_m$ at $\alpha=100\pi\omega_m$ and $G_a=20\omega_m$. Similarly, 
		the black curve represents $U=5\omega_m$ while magenta curves from dark shade to light shade represent atom-atom interactions 
		$U=6\omega_m, 7\omega_m, 8\omega_m, 9\omega_m$ and $10\omega_m$, respectively.
		The other parameters used in numerical calculation are same as in Fig.\ref{fig1}.}
	\label{fig8}
\end{figure}
If we 
increase the atomic mode coupling with intra-cavity field, atomic dressed states will absorb more phonons emitted by mechanical mirror 
of the system which will decrease the thermal excitation of mechanical mirror. Fig.\ref{fig8}(a) shows such influence of atom-field coupling 
on mechanical oscillator of the system where the effective-temperature of mirror is decreased by increasing atom-field coupling \cite{Pater06}. 
The atom-atom interactions of atomic mode will also influence the mechanical mirror similarly as SO-coupling has shown \cite{Sonam2013}. The atom-atom 
interactions also contribute to control the thermal excitation of mechanical mirror which lead to cool oscillating mirror 
to its quantum mechanical ground state, as shown in Fig.\ref{fig8}(b).

\end{document}